\documentstyle[12pt,prd,aps,epsf]{revtex}
\begin{document}
\baselineskip=20pt \draft
\title{Neutrino spin-flip effects in active galactic nuclei}
\author{M. Anwar Mughal}
\address{Department of Physics, Quaid-i-Azam University, Islamabad 45320,
 Pakistan}
\author{H. Athar\footnote{E-mail: athar@phys.cts.nthu.edu.tw}}
\address{Physics Division, National Center for Theoretical
         Sciences,
         Hsinchu 300, Taiwan \\
         and Institute of Physics, National Chiao-Tung University,
         Hsinchu 300, Taiwan}
\date{\today}
\maketitle
\begin{abstract}
\tightenlines

    We study the effects of neutrino spin-flip in the magnetic
field, $B_{AGN}$, of active
galactic nuclei (AGN) for high-energy  neutrinos ($E\geq 10^{6}$ GeV)
 originating from AGN induced
by an interplay of the violation of equivalence principle
parameterized by $\Delta f$ and the
twist in $B_{AGN}$.  We point out that a
 conversion effect may exist for
$\Delta f \sim 10^{-34}(\delta m^{2}/10^{-5} \mbox{eV}^{2})$
 independent of
gravity mixing angle. Observational consequences for this conversion effect
 are discussed.

\end{abstract}

\section{Introduction}

    In this paper, we study the spin-flip effects for high-energy
neutrinos ($E\geq 10^{6}$ GeV) originating from active galactic nuclei (AGN)
 induced by the violation of equivalence principle (VEP) and/or the magnetic
field twist as AGNs are presently considered to be a likely source of
high-energy  neutrinos \cite{ss}.  The VEP arises as different flavors
of neutrinos may couple differently to
gravity \cite{G,L,Fogli}. This essentially results from the realization that
flavor  eigenstates of neutrinos may be the
admixture of the gravity eigenstates of neutrinos with different
gravitational couplings. A magnetic field twist occurs  when
the direction of the
magnetic strength lines in the plane transverse to the neutrino
 momentum originating from AGN may not be
fixed. Several general descriptions of the
 possible effects of magnetic field twist are available \cite{general},
 as well as related
to Sun \cite{sun}, Supernovae \cite{supernova} and the early Universe
 \cite{universe}.

    The present study is particularly welcome as the new under ice or
 water
\v{C}erenkov light  detector arrays  namely AMANDA, Baikal (as well as NESTOR
 and  ANTARES), commonly known as high-energy neutrino telescopes,
 based on muon detection will
have not only the energy, angle and flavor  resolutions but also possibly the
particle and antiparticle resolution
in the electron neutrino channel near the Glashow resonance energy,
 $E\sim 6.4\cdot 10^{6}$ GeV \cite{shelly,john,lin,C}. These characteristics
make these neutrino telescopes especially suitable for the study of
 high-energy  neutrino conversions.

    We study here the spin-flip effect for Majorana type neutrinos
 in the
vicinity of the cores of active galaxies which we hereafter refer to
 as AGNs.
 Some AGNs give off a jet of matter that stream out from the core
 in a
 transverse plane and produces hot spots when the jet  strikes the
 surrounding
 matter at its other ends. For a discussion of neutrino spin-flip in
 jets
and hot spots, see \cite{DKD}.
    Previously, the spin-flip effects for AGN neutrinos
 due
to VEP
are studied in \cite{R,sahu}. The VEP is parameterized by a
 dimensionless
parameter $\Delta f$. In \cite{R}, by demanding an adiabatic
 conversion to
occur, a lower bound on neutrino
magnetic
moment $\mu$ was obtained in terms of $\Delta f$,
whereas in \cite{sahu}, the effect of possible random
fluctuation in the magnetic field of AGN,
 $B_{AGN}$ on neutrino spin precession is considered. In \cite{W},
neutrino spin-flip in
AGN due to gravitational effects (not due to VEP) and due to the presence
of a magnetic field is studied. Here we address two
 aspects of spin-flip for  high-energy  neutrinos originating
 from AGNs, viz, the spin (flavor)-precession
 with (or without) VEP and the twist in  $B_{AGN}$;
  and the adiabatic/nonadiabatic conversion due to
 an interplay of  twist in $B_{AGN}$ and the VEP. We point out that,
for latter  type of conversion effect,
 a $\Delta f$ of the order of $10^{-39}-10^{-29}$ depending on
$\delta m^{2}$ gives
 reasonably large conversion probabilities. In particular,
 we point out that the neutrino spin-flip in AGN induced by an interplay of
 VEP and  twist in $B_{AGN}$
 may give rise to {\em changes} in particle/antiparticle ratio as compared to
 no spin-flip situation in  electron neutrino channel near
 the Glashow resonance energy.

The plan of the rest of the paper is as follows. In section  II, we briefly
 discuss a matter density and a magnetic field
profile in AGN. In section  III, we discuss the
spin (flavor)-precession due to VEP and determine the value of
 $\Delta f$
needed to have the precession  probability greater than 1/2. In
 the same section, we consider in some
detail the adiabatic and non adiabatic  conversions induced by an
interplay
 of a
conceivable twist in
$B_{AGN}$, and the VEP and estimate the resulting neutrino spin
 (flavor)-conversion
 probabilities. In section  IV, we discuss a possible
 observational consequence of neutrino spin-flip in AGN
 and contrast it with the pure vacuum flavor oscillations. Finally in
 section  V, we summarize our results.

\section{The matter density and magnetic field in AGN}

Neutrino spin-precession in the context of the Sun was discussed
 in \cite{V}.
It was pointed out that the  matter effects tend to suppress the
neutrino spin-precession effect. As shown below, for AGN,  matter effects
arising due to coherent forward scattering of neutrinos off the
 matter particle background
are negligible\footnote{\tightenlines{Similar estimate for other astrophysical
 systems
like Sun and Supernovae shows that the matter effects are
 indeed
non negligible in most part of these systems \cite{luna}.}}.
    The essential conditions needed for appreciable
 spin-precession are:
 i) $\mu B \Delta r\, {\buildrel > \over {_{\sim}}} 1$, i.e., $B$ must be
 large
enough in the region of width $\Delta r$; ii) the smallness of the
 matter effects, so that neutrino spin-precession is not
 suppressed
 (see below);
and iii) there should be no reverse spin-precession of neutrinos
 on their way
to earth.
 As for the third essential condition the typical observed
 intergalactic magnetic
field for the nearby
 galaxies is
estimated to be $\sim \, O(10^{-9})$ G at a scale of Mpc, where 1 pc $\sim
 3\cdot 10^{18}$ cm \cite{book}. Taking
 a typical distance between the
earth and the AGN as $\sim \, O(10^{2})$ Mpc, we note that the effect
 induced by
intergalactic and galactic magnetic field is quite small as the
 galactic
 magnetic field is $\sim \, O(10^{-6})$ G,
thus causing negligible reverse neutrino spin-precession.

According to \cite{S}, the matter density in the vicinity of AGN
has the following profile:
$\rho (x)\, =\, \rho_{0}f(x)$ where $\rho_{0} \simeq
 1.4\cdot 10^{-13}$
g/cm$^{3}$ and $f(x)\, \simeq \, x^{-2.5}(1-0.1 x^{0.31})^{-1}$
as we take the AGN photon luminosity to be $10^{46}$ erg/s with $x\,
\equiv r/R_{S}$, $R_{S}$ being the Schwarzchild radius of AGN:
 $R_{S}\,
\simeq 3\cdot 10^{13}\left( \frac{M_{AGN}}{10^{8}M_{\odot}} \right)$ cm. We
 take the distance traversed by the neutrinos to be
$10 < \, x\, < \, 100$ in the vicinity of AGN. These imply that
the width
of the matter traversed by  neutrinos in the vicinity of the AGN is
$l_{AGN}\, \sim \, (10^{-2}-10^{-1})$ g/cm$^{2}$. In the presence of
 matter,
the effective width of
matter needed for appreciable neutrino spin-flip, on the other hand, is
$l_{\mbox{o}}\, \equiv \sqrt{2}\pi m_{N}G^{-1}_{F} \, \sim \, 2
\cdot 10^{9}$ g/cm$^{2}\, \gg l_{AGN}$.
Hence, from now on, we ignore the matter effects.

We consider now the magnetic field in the vicinity of AGN with the
following profile \cite{S}
%
%
\begin{equation}
 B_{AGN}(x)\, =\, B_{0}g(x),
\end{equation}
where $B_{0}\, \sim 5.5 \cdot 10^{4}$ G and
$g(x)=x^{-1.75}(1-0.1x^{0.31})^{-0.5}$ for $10\, <\, x\,<\, 100$.
 We will use this $B_{AGN}$ in our estimates as an example.

\section{Neutrino spin-flip due to VEP and twist in $B_{AGN}$}

The evolution equation for the two
neutrino state for vanishing gravity and vacuum mixing angles
 may be written in a frame rotating with the magnetic field as \cite{Ak}
%
%
\begin{equation}
 \dot{\iota}\dot{\psi} = H_{eff}\psi ,
\end{equation}
where $\psi^{T}=(\nu_{e}, \bar{\nu}_{\alpha})$ and $H_{eff}$ is a 2$\times $2
 matrix with $H_{11}=0$, $H_{12}=H_{21}=\mu B$ and
 $H_{22}=\delta -V_{G}+\dot{\phi}$. Here
 $\alpha\, =\, \mu $ or $\tau$, and $\dot{\phi}
 \equiv \mbox{d}\phi/\mbox{d}r$
defines the direction of rotation
of $B_{AGN}(\equiv B)$ in the plane orthogonal to the neutrino momentum.
 $\delta \, =\, \delta m^{2}/2E$, where
$\delta m^{2}\, =\, m^{2}_{j}-m^{2}_{1}\,
>\, 0$
with $E$ being the neutrino energy and $j=$ 2 or 3.
 In Eq. (2), for latter convenience, we have subtracted from the lower
diagonal
element, the upper diagonal element of the effective Hamiltonian in
 order
to make
the upper diagonal element equal to zero. This is equivalent to the
renormalization
 of the two neutrino wave functions by the same factor, which does
 not
 change the relevant precession (conversion)/survival probabilities
 \cite{Ak}.
 $V_{G}$ is the effective potential felt by the neutrinos at a
 distance $r$
from a gravitational source of mass $M$ due to VEP and in its rather
 simpler form is  given by \cite{G}
%
%
\begin{equation}
 V_{G}\, \equiv \, \Delta f\beta (r)E,
\end{equation}
where $\Delta f \, =\, (f_{\alpha}-f_{1})(f_{\alpha}+f_{1})^{-1}$ and
$\beta (r)\, =\, G_{N}Mr^{-1}$
is the gravitational potential in the Keplerian approximation,
with $G_{N}$ being the gravitational
constant. Here $f_{1}G_{N}$ and $f_{\alpha}G_{N}$ are the gravitational
 couplings
 respectively for $\nu_{e}$ and $\bar{\nu}_{\alpha}$, such that
 $f_{1}\,
 \neq \, f_{\alpha}$. Let us note that in the vicinity of AGN,
the $V_{G}$ due to AGN dominates \cite{japan}.

 We consider mainly the following two neutrino flavors: $\nu_{e}$
 and $\nu_{\tau}$ in the subsequent discussion in this section,
 motivated by the fact that
 the initial fluxes of these neutrino states are estimated to be
maximally asymmetric, typically with
 $(\nu_{\tau}+\bar{\nu}_{\tau})/(\nu_{e}+\bar{\nu}_{e})\,
 {\buildrel < \over {_{\sim }}} \, 10^{-5} $,
 according to various models of AGN \cite{MPR}. Presently, the
 high-energy
 neutrino flux from AGNs can dominate over the atmospheric neutrino background
 typically for
$E \geq 10^{6}$ GeV. The current empirical upper bounds on
high-energy neutrino flux, for instance, from AMANDA (B10), is
relevant typically for $E \, \leq  10^{6}$ GeV \cite{nu}. Let us
mention here that the upper bound discussed in \cite{wb} does not
apply to (diffuse) high-energy  neutrino flux originating from
cores of AGNs because these sources do not contribute dominantly
to the observed ultrahigh-energy cosmic ray flux.

 In this section, we intend to discuss in some detail the possible
 effects arising due to
 interaction of neutrino magnetic moment, $\mu $, with $B_{AGN}$,
 to
 enhance this ratio, that is, to obtain
$(\nu_{\tau}+\bar{\nu}_{\tau})/(\nu_{e}+\bar{\nu}_{e})\, \gg \, 10^{-5} $.
 In this context, we now propose to study  the various main
 possibilities arising from the
relative comparison
between $\delta$, $V_{G}$ and $\dot{\phi}$ in Eq. (2).

{\noindent \bf Case 1.} $V_{G}\, =\, \dot{\phi}\, =0$. For
 constant $B$,
we obtain the following expression
for spin-flavor precession probability
 $P(\nu_{e}\rightarrow \bar{\nu}_{\alpha})$ by solving Eq. (2):
%
%
\begin{equation}
 P(\nu_{e}\rightarrow \bar{\nu}_{\alpha})\, =\,
 \left[\frac{(2\mu B)^{2}}{(2\mu B)^{2}+X^{2}}\right]
 \sin^{2}\left(\sqrt{(2 \mu B)^{2}+X^{2}}\cdot \frac{\Delta r}{2}\right),
\end{equation}
with $X\, =\, \delta$. We now discuss the relative comparison between
 $2\mu B$
and $\delta $ and evaluate
$P$
for corresponding $\delta m^{2}$ range.

{\noindent a)} $\delta \ll 2\mu B$. Using $B$ given in Eq. (1) for
$\mu \, \sim \, 10^{-12}\, \mu_{B}$
\cite{F}, the condition $\delta \ll 2\mu B$ implies
 $\delta m^{2}\, \ll \, 5\cdot 10^{-4}$ eV$^{2}$ with
$E\, \sim \, 10^{6} $ GeV. We take here $\delta m^{2}\, \sim \, 5\cdot 10^{-6}$
 eV$^{2}$ as an example.  The expression (4) for $P$ then reduces to
%
%
\begin{equation}
 P(\nu_{e}\rightarrow \bar{\nu}_{\alpha})\, \simeq \, \sin^{2}
 (\mu B \Delta r).
\end{equation}
The phase of $P$ can be of the order of unity if $\mu B\Delta r \, =\,
\frac{\pi}{2}$ (or if $\mu B\Delta r\, {\buildrel > \over {_{\sim}}} 1$) for a
constant $B$.
Evidently, this $P$ is independent of $E$.  According to Eq. (1), the
$B_{AGN}$ varies with distance so that to have maximal depth of spin-flavor
precession, we need
to integrate the strength of the magnetic field along the neutrino trajectory.
Thus, for maximal depth of $\nu_{e}\rightarrow \bar{\nu}_{\alpha}$
 precession,
we require in Eq. (5) that
%
%
\begin{equation}
 \mu \int^{r^{\prime}}_{0} {\mbox d}r^{\prime} B(r^{\prime}) \,
 {\buildrel > \over {_{\sim}}} \, 1.
\end{equation}
We note that Eq. (5) [along with Eq. (6)] give
$P(\nu_{e}\rightarrow
 \bar{\nu}_{\alpha})\, > \, 1/2$  for the $B_{AGN}$ profile given by Eq. (1)
with
$\mu \, \sim \,
10^{-12}\, \mu_{B}$.
Thus, an energy independent permutation (exchange) between $\nu_{e}$
and $\bar{\nu}_{\alpha}$ may result with $P>\, 1/2$.
 This energy {\em independent} permutation of energy spectra of $\nu_{e}$
 and
$\bar{\nu}_{\alpha}$ for
small $\delta m^2$  follows from the fact that Eq. (5) also gives
$P(\bar{\nu}_{\alpha}\, \rightarrow \,
 \nu_{e}$) since we are considering a two neutrino state system.
For another magnetic field strength profile of
 AGN \cite{B} we obtain
 $P(\nu_{e} \rightarrow \bar{\nu}_{\alpha}) \, > 1/2$ for $\mu \, \sim \,
 2\cdot 10^{-16}\mu_{B}$  [this profile suggests a
 constant magnetic field
$\sim O(10^{4})$ G for $ x \, {\buildrel > \over {_{\sim}}} \, 10$].
 We thus obtain the same $P$ value ($P \, >\, 1/2$) with a 4 orders of
magnitude small $\mu $ for this $B_{AGN}$ profile for same $\delta m^2$.
 Therefore, if $\mu$ turns out to be $\sim \, O(10^{-16})\, \mu_{B}$
and if empirically it is found that, for instance,
 $P(\nu_{e}\rightarrow \bar{\nu}_{\tau})\, >\, 1/2$ for small
$\delta m^{2}$ then this situation may be an evidence  for the latter
 $B_{AGN}$
profile.
Let us further note that this small value of
$\delta m^{2}$ ($\delta m^{2}\, \sim \, 5\cdot 10^{-6}$ eV$^{2}$)
 is not only
interesting in the context of Sun \cite{sun} but also Supernovae
 \cite{supernova}.\\
{\noindent b)} $\delta \, \simeq \, 2\mu B$. Here $\delta m^{2}$
corresponds to $5\cdot 10^{-4}$ eV$^{2}$. In
this case expression (4) for $P$ reduces to
%
%
\begin{equation}
 P(\nu_{e}\rightarrow \bar{\nu}_{\alpha})\, \simeq \, 1/2\,
 \sin^{2}(\sqrt{2}\mu B\Delta r).
\end{equation}
Thus, for $\delta m^{2}\, \simeq \, 5\cdot 10^{-4}$ eV$^{2}$, energy
dependent distortions may
result in {\em survived} and {\em precessed} neutrino energy spectra
with $P\, {\buildrel < \over {_{\sim}}} \, 1/2$.\\
{\noindent c)} $\delta \gg \, 2\mu B$, that is,
$\delta m^{2}\, \gg \, 5\cdot 10^{-4}$ eV$^{2}$. Energy
dependent distortions may result for relatively large
$\delta m^{2}$ with $P\, <\, 1/2$. For instance, consider
$\delta m^{2} \, \sim \, 10^{-3}$ eV$^{2}$ relevant for
 atmospheric
neutrino problem \cite{YU}.
 The $(\nu_{e}+\bar{\nu}_{e})/ (\nu_{\tau}+\bar{\nu}_{\tau})$ ratio as well as
 $(\nu_{e}+\bar{\nu}_{e})/ (\nu_{\mu}+\bar{\nu}_{\mu})$
 will have energy dependence in this case.
 Among the $\nu_{\mu}$ and $\nu_{\tau}$ channels, the
 spin-flavor precessions
lead to an energy dependent $(\nu_{\tau}+\bar{\nu}_{\tau})/
 (\nu_{\mu}+\bar{\nu}_{\mu})$.
 This situation may be realized by replacing $\nu_{e}$ by $\nu_{\mu}$
 and $\alpha $ by $\tau$ in Eq.
(2) with the corresponding changes in $\delta m^{2}$ and in $V_{G}$.
 For comparison, let us note that the pure
vacuum flavor oscillations lead to an energy independent ratio equal to 1/2,
 that is, $(\nu_{\tau}+\bar{\nu}_{\tau})/ (\nu_{\mu}+\bar{\nu}_{\mu})\, \sim
 1/2$
 \cite{AJY}. Therefore, an energy dependent ratio different from 1/2 may
provide an evidence for high-energy neutrino spin-flip in AGN.
  It is relevant here to mention that the
 future/existing  high-energy neutrino telescopes may attempt to
 measure the three ratios
 $(\nu_{\tau}+\bar{\nu}_{\tau})/(\nu_{\mu}+\bar{\nu}_{\mu})$,
 $(\nu_{e}+\bar{\nu}_{e})/(\nu_{\mu}+\bar{\nu}_{\mu})$
 as well as $(\nu_{\tau}+\bar{\nu}_{\tau})/(\nu_{e}+\bar{\nu}_{e})$ of the
 absolute fluxes of high-energy  neutrinos
 and possibly the energy dependence in this ratio
 [see section  IV for some further discussion].

Let us note that all these spin (flavor)-precession situations are
realized without VEP and
magnetic field twist in $B_{AGN}$ as a  spin (flavor)-precession for
 AGN neutrinos may develop
even without VEP and gravitational neutrino flavor dependent effects.
 Thus, the cause of change (as compared to no
precession situation) in the ratios of the $(\nu_{\tau}+\bar{\nu}_{\tau})$,
 $(\nu_{\mu}+\bar{\nu}_{\mu})$ and
$(\nu_{e}+\bar{\nu}_{e})$ fluxes, as well as an energy
dependence in these ratios, in future/existing high-energy neutrino
telescopes may not only be attributed to
VEP and/or gravitational effects depending on relevant $\delta m^{2}$ range.

{\noindent \bf Case 2.} $V_{G}\, =\, 0, \, \dot{\phi}\neq 0$. For constant
 $B$
and $\dot{\phi}$, we obtain
the  expression for precession  probability (for small $\delta $) by
substituting $\dot{\phi}$ for $X$ in
Eq. (4).
 We first take $\dot{\phi}\, \sim \, 2\mu B$, thus $\delta \, \ll \,
\dot{\phi}$
for $\delta m^{2}\, \sim \,
5\cdot 10^{-6}$ eV$^{2}$ [as considered in case 1a)].
 Note that in this expression for precession
probability, the   sign of $\dot{\phi}$ is
unimportant. It is natural
to
suggest that the total rotation angle of the AGN magnetic field is
 restricted
by $\Delta \phi \, {\buildrel < \over {_{\sim}}}
 \pi$. Thus, for instance, a twist appears, when high-energy neutrinos
 cross the toroidal magnetic
field with magnetic strength lines winding around the spherically
accreting matter disk in AGN.
 In this case the maximal rotation angle is $\pi$,
i.e., the above bound is satisfied.
    The field twist can be characterized by the scale of the twist,
$r_{\phi}$, such that  $r_{\phi}\equiv \pi/\dot{\phi}$, so that on the
 way,
the total rotation angle (for uniform rotation), equals to
 $\Delta \phi \, =\,
\pi$. Let us define the critical rotation scale as
 $r^{c}_{\phi}\, \equiv \,
\pi/2\mu B$ \cite{Sm}. Note that this $r^{c}_{\phi}$ coincides with the
precession length $l_{\mbox{p}}\,
 [\equiv \, (2\mu B)^{-1}]$ apart from  a factor of $\pi$ and on dimensional
 grounds is the simplest possibility.
  For appreciable magnetic field
twist effects, evidently we require $r_{\phi}\, {\buildrel < \over
{_{\sim}}} r^{c}_{\phi}$. Comparing $r^{c}_{\phi}$ with the
distance from the center of AGN in units of $x$ (or $R_{S}$), we
find that $r \sim r^{c}_{\phi}\,$ for
 a  $B$ that is smaller than the
available $B_{AGN}$ given by Eq. (1). In this case
 $P(\nu_{e}\rightarrow \bar{\nu}_{\alpha})$
reduces to Eq. (7). Thus, for
small $\delta m^{2}$, we obtain
here $P\, {\buildrel < \over {_{\sim}}} \, 1/2$. This case  can
 therefore be
differentiated from the previous one by concentrating on
$P$ value. For small $\delta m^{2}$ [case 1a)] previously we have
 $P\, >\,
1/2$.
The magnetic field twist effects
here may
give rise to energy independent spin (flavor)-precession between
 $\nu_{e}$ and $\bar{\nu}_{\tau}$. However, here unlike previous
 case for
small
$\delta m^{2}$, the required
$B$ has an upper bound for a naturally scaled field twist. For
$\dot{\phi}\, \ll \, 2\mu B$, we obtain case 1a)
 whereas for $\dot{\phi}\, \gg \, 2\mu B$, we obtain case 1c).

 For $\delta \, \sim \, -\dot{\phi}$, the  spin (flavor)-precession
 results
from a
cancellation between $\delta $ and $\dot{\phi}$ which for a naturally
scaled $\dot{\phi}$ corresponds
to $\delta m^{2}\, {\buildrel < \over {_{\sim}}} \, 5\cdot 10^{-4}$
 eV$^{2}$
with $P\, >\, 1/2$, whereas the opposite sign of
$\dot{\phi}$ results in
suppression of $P$. Thus, for large $\delta m^{2}$ (but comparable to
$\dot{\phi}$),  energy
 dependent distortions may occur with $P\, >\,  1/2$. For
$\delta \, \gg \, \dot{\phi}$, this case reduces
to case 1c).

{\noindent \bf Case 3.} $V_{G}\neq \, 0, \, \dot{\phi}\, =\, 0$
(with small $\delta$, that is, $\delta \, \ll \,
2\mu B$). For constant $V_{G}$
and $B$, we obtain from Eq. (4) the relevant precession
 probability
expression by replacing $X$
with $V_{G}$.
 If $V_{G}\, \ll \, 2\mu B$ then using Eq. (1) and Eq. (3), we obtain
 $\Delta f \, \ll \, 6\cdot 10^{-32}$. We take here
$|\Delta f|\, {\buildrel < \over {_{\sim}}} \, 10^{-34}$ as our criteria
and so consequently the corresponding  $P$ reduces to (5). This results
 in
$P\, >\, 1/2$ with no energy dependence. Thus this case coincides with
 case
1a) for small $\Delta f\, ({\buildrel < \over {_{\sim}}} \, 10^{-34})$
depending on the given $B_{AGN}$ profile.
     Consequently, if there is a VEP at the level of $10^{-34}$ or less, a
 spin (flavor)-precession for  neutrinos  may occur in the
vicinity of AGN with small $\delta m^{2}$. Evidently, this value
of $\Delta f$ is independent of the gravity mixing angle \cite{G}.
Let us note in passing that this value of $\Delta f$ is (much)
lower than the one obtained in
 \cite{Fogli}. For
$\Delta f\, {\buildrel > \over {_{\sim}}} \, 10^{-34}$, energy
 dependence in
$P$ results with $P\, {\buildrel < \over {_{\sim}}} \, 1/2$. For large
$\delta $ ($\delta \, \simeq \, V_{G}$) see case 5 and if
$\delta \, \gg \, V_{G}$ then this case reduces to 1c).
    The upper bound for $\Delta f$ obtained in this case has only a
linear
energy dependence,
 whereas the other necessary requirement [Eq. (6)] does not depend on $E$
 for
 small $\delta$.
 This is in sharp contrast to the situation discussed in case 5,
where both the level crossing as well as the adiabaticity conditions
 depend
on $E$. Thus, to summarize, we have pointed out in this case that for
high-energy  neutrinos originating from AGN, a  spin (flavor)-precession
 may
develop in the vicinity of AGN if
$\Delta f \, {\buildrel < \over {_{\sim}}} \, 10^{-34}$ yielding, for
instance,
 $(\nu_{\tau}+\bar{\nu}_{\tau})/(\nu_{e}+\bar{\nu}_{e})\, \gg \, 10^{-5} $.

The observational consequences of the high-energy  neutrino spin-flavor
precessions discussed in the previous three cases are the energy dependence in
the relevant ratio of the fluxes as well as a possible change in the ratios
 with respect to
the pure vacuum flavor oscillations.
 With the improved information on the relevant
neutrino mixing parameters, these cases may in principle be disentangled from
 each other.

{\noindent \bf Case 4.} $V_{G}\, =\, \dot{\phi}$ (for small $\delta$).
This results in conversion
effect in contrast
to the previously considered three cases [which are  spin
(flavor)-precession effects].

Two conditions are
essential for an adiabatic conversion: i) level crossing and ii)
 adiabaticity.
    The level crossing is obtained by equating the diagonal element
 of
the effective Hamiltonian in Eq. (2), i.e., $V_{G}\, =\, \dot{\phi}$
implying $\Delta f\, \propto \, E^{-1}$ (or a {\em linear} dependence
 of
$\dot{\phi}$ on $E$). For  $\bar{\nu}_{e}\rightarrow
\nu_{\alpha}$ conversions, if $\Delta f\, >\, 0$ (both for particles
 and
antiparticles)
then there is no level crossing as $\dot{\phi}$  is negative for this
channel.
 If $\Delta f\, <\, 0$, then the level crossing shifts to
antiparticle channel ($\bar{\nu}_{e}\rightarrow \nu_{\alpha}$).
Thus, a simultaneous deficit/enhancement in both $\nu_{e}$
and $\bar{\nu}_{e}$ spectra (and in  $\nu_{\tau}$ and
$\bar{\nu}_{\tau}$ spectra) is not expected
 due to an interplay of VEP and the twist in $B_{AGN}$ unless $\Delta f$ has
 different sign  for particles and antiparticles.
 The level crossing is induced by a naturally
scaled field twist for $\Delta f \, {\buildrel < \over {_{\sim}}}  \,
10^{-34}$ ,
 that is, when $r^{c}_{\phi}/r\, {\buildrel > \over {_{\sim}}} \, 1$ (see
 case
2
also).
    Let us note that this level crossing is induced by an
 interplay of magnetic field twist
and VEP for  neutrinos with small $\delta m^{2}$
 ($\delta m^{2}< 5\cdot 10^{-6}$ eV$^{2}$). This is a  characteristically
distinct feature of a more realistic situation of having magnetic strength
lines  winding
around the nearly spherical matter disk.
However, level crossing alone is not a sufficient condition for a
complete conversion. As stated earlier,
 adiabaticity is the other necessary condition that determines
the extent of conversion. If there is only level crossing and no
 adiabaticity
 at the level crossing
then there is no conversion of  electron neutrinos into anti tau
neutrinos. In the remaining part
 of this case, we discuss quantitatively the latter condition, that is,
 the adiabaticity.

    The adiabaticity condition assumes the slowness of variation in
$V_{G}$ and is given by \cite{lim}:
%
%
\begin{equation}
 \kappa_{R}\, =\, \frac{2(2\mu B)^{2}}{|\dot{V}_{G}|}.
\end{equation}
    This is the adiabaticity parameter in the resonance for uniform
 magnetic field twist ($\ddot{\phi} = 0$).
 A conversion is adiabatic if $\kappa_{R}
\, {\buildrel > \over {_{\sim}}} \,  1$. Notice that here
$\kappa_{R}\, \propto \, E^{-1}$. Since $\kappa_{R}$ depends
 (quadratically) on
$B$, thus adiabaticity of conversion is essentially
determined and controlled by the given $B$ profile. By requiring an adiabatic
conversion to occur, we can
obtain
$B_{\mbox{ad}}$ from Eq. (8).
Using Eq. (1) and for $\Delta f\, \sim \, 10^{-34}$
(a requirement of level crossing), we obtain $B_{\mbox{ad}}\, < \,
B_{AGN}$. It is interesting to note that the $B_{\mbox{ad}}$
 does not depend on
 any $B$ profile of AGN, it is determined rather by the gradient of
 $V_{G}$. Thus,
an adiabatic conversion may occur for $\Delta f\, \sim \, 10^{-34}$ or
less depending on $\delta m^{2}$ in a
uniform magnetic field twist. Let us emphasize that this adiabatic level
 crossing
is induced
 by the change in the gravitational potential rather than the change in
 effective matter density. A general expression for neutrino spin-flavor
conversion probability including the effect of non adiabaticity
 ($\kappa_{R}\, < 1$), using Eq. (2) is \cite{kim}
%
%
\begin{equation}
 P(\bar{\nu}_{e}\rightarrow \nu_{\tau}) =
 \frac{1}{2}-\left\{\frac{1}{2}-\exp\left(-\frac{\pi}{2}\kappa_{R}\right)
 \right\}\cos 2\theta_{B_{i}}\cos 2\theta_{B_{f}},
\end{equation}
where $\tan 2\theta_{B_{i}}=(2\mu B)/(\delta -V_{G})$ is evaluated at the
high-energy neutrino production site in the vicinity of AGN and
$\tan 2\theta_{B_{f}}=(2\mu B)/(\delta -V_{G})$ is evaluated at the exit.
 In Fig. 1,
 we display $P$ using Eq. (9) for some representative values of $\Delta f$
 with $\delta m^{2}\, \sim 10^{-10}$ as a function of $E$ for
 illustrative purpose only.
 From Fig. 1, we notice that for $E\sim 6.4\cdot 10^{6}$ GeV, the $P$ is
rather large ($\sim 0.6-0.7$), thus leading to a suppression in the
$\bar{\nu}_{e}$ flux.

    Nonuniform field twist ($\ddot{\phi}\, \neq \, 0$) changes the
adiabaticity condition (8). It now reads
%
%
\begin{equation}
 \kappa_{\phi}\, =\, \frac{2(2\mu B)^{2}}{|\dot{V}_{G}-\ddot{\phi}|}.
\end{equation}
    Thus, for $\ddot{\phi}\, \simeq \dot{V}_{G}$, we may have a large
enhancement in $\kappa_{\phi}$. For a naturally scaled field twist, the
 total
rotation angle for a nonuniform magnetic field twist is given by
\cite{supernova}
%
%
\begin{equation}
 \Delta \phi \, \sim \, \kappa^{-1}_{R},
\end{equation}
i.e., the total rotation angle is given by the inverse of the adiabaticity
parameter for a uniform  magnetic field twist.
 Clearly, only modest improvement
in $\kappa
_{R}$ may be achieved for a naturally scaled magnetic field twist.
The corresponding conversion probability $P$ in this case is energy
 dependent.
 Thus, observationally, we may obtain here
$(\nu_{\tau}+\bar{\nu}_{\tau})\, \sim \, (\nu_{e}+\bar{\nu}_{e})$,
  due to an
 adiabatic conversion induced by an interplay of $\dot{\phi}$ and $V_{G}$
 in the vicinity of the AGN.
 For large $\delta $, comparable to
$V_{G}$ and $\dot{\phi}$, see  case 6.

{\noindent \bf Case 5.} $V_{G}\, \simeq \, \delta, \, \dot{\phi}\, =\, 0$.
This situation also results
in conversion effects (as opposed to cases 1-3), however see case 3 also.
The level crossing implies
%
%
\begin{equation}
  \Delta f \, \simeq
 2 \cdot 10^{-34}\left(\frac{\delta m^{2}}{10^{-5}eV^{2}}\right).
\end{equation}
Note that relative sign between $\delta $ and $V_{G}$ is important for level
crossing.

    If $\Delta f\, >\, 0$, both for particles and antiparticles, then
both $\nu_{e}$ and $\bar{\nu}_{e}$ will transform simultaneously, whereas if
 $\Delta f\, <\, 0$, both for particles and antiparticles, no level crossing
takes place. On the other hand, if $\Delta f$ changes sign for particles and
antiparticles, level crossing between particles {\em or} antiparticles will
take place. Thus, this case can be distinguished from the previous case.

 It is important to note that from the level crossing it follows that
$\Delta f\, \propto \, E^{-2}$, i.e.,
an inverse quadratic $E$ dependence on $\Delta f$. Thus, the level
crossing induced by the VEP alone has a
{\em different} energy dependence on $\Delta f$ as compared to the level
crossing induced by an interplay of $\dot{\phi}$
 and the VEP (see previous case). The relevant adiabaticity condition
 may be
written as
%
%
\begin{equation}
 B_{\mbox{ad}}\,
 {\buildrel > \over {_{\sim}}} \, 3\cdot 10^{2}\,
 \mbox{G}\left(\frac{10^{-12}\mu_{B}}{\mu}\right)\left(\frac{\Delta
 f}{10^{-29}}\right)^{\frac{1}{2}}\left(\frac{10R_{S}}{r}\right).
\end{equation}
We note that $B_{\mbox{ad}}\, {\buildrel < \over {_{\sim}}}
 \, B_{AGN}$ for 10$\, <\, x\, <\, $100. The adiabaticity parameter here
has the {\em same} energy
dependence on $E$ as in case 4. Thus, the adiabatic conversion may occur
giving rise to energy
dependent distortions with corresponding conversion probability greater
than 1/2. For large $\delta $
whereas a  spin (flavor)-precession is suppressed [see case
1b) and 1c)], an adiabatic conversion may result with $P\, >\, 1/2$ for
large $\Delta f$ thus resulting in correspondingly different
 observational consequences. For $\delta \, \ll \,
V_{G}$ this case reduces to case 3 whereas for $\delta \, \gg \, V_{G}$,
 we
obtain case 1c).

    It follows from the discussion in cases 4 and 5 that a
{\em nonzero} $\Delta f$ is needed to induce an adiabatic level crossing
 with $P\, >\, 1/2$. It is in contrast to cases 1, 2  and 3 where a
 spin-flip
may occur through spin (flavor)-precession without $\Delta f$
 with $P\, >\, 1/2$.

{\noindent \bf Case 6.} If $\delta $, $V_{G}$ and $\dot{\phi}$ are
of the same order of magnitude then we
have two possibilities: the $V_{G}$ and $\dot{\phi}$ terms cancel each
 other.
Then, effectively case
1 a) is recovered.
 On the other hand, if $V_{G}$ and $\dot{\phi}$ tend to add up, then
effectively (apart from a
factor of 2) we obtain either case 2 or case 5.

    From the discussion in the previous cases, it follows that neutrino
spin-flavor precessions/conversions may occur in several situations depending
on the range of relevant neutrino mixing parameters.

\section{Possible observational consequences of neutrino spin-flip in AGN}

    In this section, we discuss in some detail the potential of the
 future high-energy neutrino telescopes to
possibly determine some observational consequences of neutrino spin-flip
in AGN through  examples.

    The planned high-energy neutrino telescopes may in
principle differentiate between the three  neutrino
flavors ($e$, $\mu $ and $\tau $) considered so far in this paper \cite{john}.
 The particular relevance here is of the electron neutrino channel, in which
 the downward going $\bar{\nu}_{e}$ interaction rate (integrated over all
angles) is estimated to be  an order of magnitude higher than that of
 $(\nu_{e}+\bar{\nu}_{e})$ per
 Megaton per year at $E\sim \, 6.4\cdot  10^{6}$ GeV \cite{GH}.
 This an order of magnitude
difference in interaction rate of downward going
$\bar{\nu}_{e}$ relative to  $(\nu_{e}+\bar{\nu}_{e})$ deep inelastic
 scatterings is due to Glashow resonance encountered by
 $\bar{\nu}_{e}$ with $E\, {\buildrel > \over {_{\sim}}} \, 10^{6}$ GeV
 when they interact with electrons near or inside the detector.
 The upward going $\bar{\nu}_{e}$, on the other hand, while passing
 through the earth, at these energies, are almost completely absorbed
 by the earth. Thus, for instance, if $E \, \sim \, 6.4\cdot  10^{6}$ GeV,
an energy
resolution $\Delta E/E\, \sim \, 2\Gamma_{W}/M_{W}\, \sim \, 1/20$,
where $\Gamma_{W}\, \sim $ 2 GeV is the
width of Glashow resonance and $M_{W}\, \sim \, $80 GeV, may be needed
 to empirically differentiate between $\nu_{e}$ and $(\nu_{e}+\bar{\nu}_{e})$.
   The $\bar{\nu}_{e}$ and $(\nu_{e}+\bar{\nu}_{e})$ essentially produce a
single shower event. Thus, the planned high-energy neutrino
telescopes may in principle attempt to measure the
$\nu_{e}/\bar{\nu}_{e}$ ratio near the Glashow resonance energy in
addition to
 identifying
 ($\nu_{\tau}+\bar{\nu}_{\tau}$), ($\nu_{\mu}+\bar{\nu}_{\mu}$) as well as
 ($\nu_{e}+\bar{\nu}_{e}$) events  separately by measuring the ratio of these
fluxes. This may allow us to corroborate the neutrino mixing effects
 somewhat  meaningfully.

    The near future large high-energy
 neutrino telescopes may attempt to utilize this
enhancement in the $\bar{\nu}_{e}$ cross-section over electrons to measure
 the high-energy (antielectron) neutrino flux. Therefore, it is
useful to ask for what possible range of neutrino mixing parameters,
 the high-energy  $\bar{\nu}_{e}$ flux
could be suppressed (or enhanced). In the remaining part of this section,
 we elaborate such a possibility. Let us remark here that at present the
 absolute normalization of the high-energy neutrino flux is basically
 unknown \cite{debate}.
The suppression or enhancement for high-energy
 $\bar{\nu}_{e}$ flux correlated to the direction of the source alongwith the
corresponding changes in the remaining neutrino flavors as pointed out in
this paper depends only on the neutrino mixing parameters (and on the
 source).

 Let us comment on the implications of current atmospheric and
 solar neutrino results on our analysis presented in section
 III. A recent global three neutrino oscillation study \cite{global} of neutrino data
 indicates that the best fitted $\delta m^{2}$ and
 $\sin^{2}2\theta $ values to solve the atmospheric neutrino
 problem in terms of $\nu_{\mu}\to \nu_{\tau}$ oscillations are
 typically $\sim 10^{-3}$ eV$^{2}$ and $\sim 1$. On the other hand,
 presently there exists more than one solution to solve the
 solar neutrino problem in terms of $\nu_{e}\to \nu_{\alpha}$
 oscillations. For SMA (MSW) solution, the $\delta m^{2}$ and
 $\sin^{2}2\theta $ values are $\sim 10^{-5}$ eV$^{2}$ and $\sim 10^{-2}$,
 for LMA (MSW) solution, these are $\sim 10^{-5}$ eV$^{2}$ and $\sim 1$, whereas
 for VAC solution, these are $\sim 10^{-10}$ eV$^{2}$ and $\sim
 1$, respectively. The LOW solution values are close to that
 of LMA (MSW) solution. Following \cite{bona}, and using these values of
 $\delta m^{2}$ and $\sin^{2}2\theta $, we
 note that {\em energy independent} pure vacuum neutrino flavor
 oscillations occur between the AGN and the earth irrespective of
 the specific oscillation solution for solar neutrino problem.

    In order to further contrast the spin-flip effects studied in this
 paper with the pure vacuum flavor oscillations for (downward
going) high-energy neutrinos originating
 from AGNs let us
 emphasize that vacuum flavor oscillations lead to an energy independent
same ratio for the three flavors, i.e., $F_{e}: F_{\mu}: F_{\tau} = 1: 1: 1$,
where $e\equiv (\nu_{e}+\bar{\nu}_{e})$, etc., at the level of intrinsic
 electron neutrino flux $F^{0}_{e}$.
 It is so because firstly the
matter effects are basically negligible in the vicinity of the yet known
sources of
high-energy  neutrinos as well as between the source and the earth and
secondly the sources are considered to be cosmologically distant and that the
intrinsic ratio of the high-energy  neutrinos is 1 : 2 : 0.
 Therefore, a deviation from 1 : 1 : 1 for the final ratios correlated to the
 direction of the source as well as an energy dependence may
provide an example of neutrino spin-flip effect in AGN. A simple
relevant remark is in order here. The pure vacuum flavor oscillation length
 is given
by, $l_{f}\sim 4E/\delta m^{2} $, whereas the spin-flavor
 precession length is (defined earlier as $l_{\mbox{p}}$),
$l_{sf}\sim 1/2\mu B$. For $\delta m^{2}$ range under discussion, i.e.,
$10^{-10}\leq \delta m^{2}/\mbox{eV}^{2}\leq 10^{-3}$, and for the
typical $E$ value range, i.e., $10^{6}\leq E/\mbox{GeV}\leq 10^{7}$,
with $\mu \sim 10^{-12}\mu_{B}$ and $B\equiv B_{AGN}$ given by Eq. (1), we
note that $l_{sf}<l_{f}$. Therefore, spin-flip effects may dominate in the
vicinity of the AGN. The pure vacuum flavor oscillations that may dominate
 between
the AGN and the earth are essentially an energy independent effect. Thus the
energy dependence due to neutrino spin-flip in AGN in for instance
($\nu_{e}+\bar{\nu}_{e}$)/($\nu_{\tau}+\bar{\nu}_{\tau}$) will survive
providing a signature of neutrino spin-flip in AGNs (see Fig. 1).

    To disentangle the neutrino spin-flip effects from  pure vacuum
flavor oscillation effects, a suitable energy range
 $\sim 4M_{W}\Gamma_{W}/m_{e}$ can be centered at
 $E\sim 6.4 \cdot 10^{6}$ GeV. The vacuum neutrino flavor mixing parameters
(namely $\delta m^{2}$ and $\sin^{2}2\theta $) will presumably be get
 measured in various terrestrial experiments and so the corresponding effects
for high-energy neutrinos can reliably be isolated from the spin-flip effects
discussed here.

    In case of spin-flavor precessions between
($\nu_{\mu}+\bar{\nu}_{\mu}$) and ($\nu_{\tau}+\bar{\nu}_{\tau}$), which may
happen for the range of $\delta m^{2}$ values given in case 1c) of previous
section, the observational consequence is a change in the value of
 ($\nu_{\tau}+\bar{\nu}_{\tau}$)/($\nu_{\mu}+\bar{\nu}_{\mu}$) ratio as
compared to that of pure vacuum flavor oscillations along with
possible energy dependence. The empirical distinction between
$\nu_{\mu}$ and $\bar{\nu}_{\mu}$ as well as $\nu_{\tau}$ and
$\bar{\nu}_{\tau}$ is currently not envisaged for the typical
high-energy neutrino telescopes. The spin-flavor precession
effects discussed in cases 1-3 leads to precessions of the type
$\nu_{e}\rightarrow \bar{\nu}_{\mu}, \bar{\nu}_{\tau}$ and
$\bar{\nu}_{e}
 \rightarrow \nu_{\mu}, \nu_{\tau}$ simultaneously. Thus, in this case the
ratio $\nu_{e}/\bar{\nu}_{e}$ is the same, however, energy dependence in the
 ratio $\nu_{e}/\bar{\nu}_{e}$ and a
change in the non electron neutrino flux ratios here remain a distinctive
feature of spin-flavor precessions depending on $\delta m^{2}$ values.
 The energy dependence in $\nu_{e}/\bar{\nu}_{e}$ ratio due to production
 should be essentially absent in case $\nu_{e}$ and $\bar{\nu}_{e}$ come
from the same parent particle, for instance, from $\mu$.

    There are several situations (case 4-5) as discussed in the previous
section in which $\nu_{e}\rightarrow \bar{\nu}_{\mu},\bar{\nu}_{\tau}$
 spin-flavor
conversions may occur. As pointed out earlier in this section, distinction
between $\nu_{e}$ and $\bar{\nu}_{e}$ may become possible near the Glashow
 resonance energy so this possibly
gives a better chance to identify an observational consequence of neutrino
spin-flip through spin-flavor conversions. For $\Delta f \sim 10^{-34}$, if
VEP is different for neutrinos and antineutrinos then the energy dependent
spin-flavor conversions as discussed in case 5 may give rise to change in
$\nu_{e}/\bar{\nu}_{e}$ ratio, in addition to change in ($\nu_{\tau}+
 \bar{\nu}_{\tau}$) or ($\nu_{\mu}+\bar{\nu}_{\mu}$), which ever the case may
be.  However, if VEP is the same for neutrinos and antineutrinos,
then this situation coincides with the previous situation of
spin-flavor precession, i.e., no change in the
$\nu_{e}/\bar{\nu}_{e}$ ratio. Thus, for instance, absence (or
enhancement, depending on sign of $\Delta f$) of $\bar{\nu}_{e}$
 events near Glashow energy and energy dependence
and enhancement in ratios of other neutrino flavors from an AGN may provide
 an observational consequence for
neutrino spin-flip in AGN.

    An interesting situation may arise after the incorporation of magnetic
field twist effects, as discussed in case 4 which also give rise to change in
$\nu_{e}/\bar{\nu}_{e}$ ratio but for different (small) $\delta m^{2}$ values,
irrespective of nature of VEP. The pure $\nu_{e}\rightarrow \bar{\nu}_{e}$ or
$\bar{\nu}_{e}\rightarrow \nu_{e}$ (though suppressed \cite{aps}) may also
take place
giving rise to changes in $\nu_{e}/\bar{\nu}_{e}$ ratio (for instance,
different from unity) possibly with no energy dependence or change in
($\nu_{\tau}+\bar{\nu}_{\tau}$)/($\nu_{\mu}+\bar{\nu}_{\mu}$) ratio.
 This can be realized from
the discussion in case 4 of section  III where it is pointed out
 that an
interplay between VEP and a naturally scaled
field twist leads to conversions in either
 $\nu_{e}$ or $\bar{\nu}_{e}$
channel but not in both channels
simultaneously. Note that in this case nonzero $\Delta f$
 and a  nonzero $\dot{\phi}$ is needed.

    A relevant remark is that ``matter like" effects induced by the
presence of nonzero $\Delta f$ (along with nonzero $\dot{\phi}$)
differentiates  between particles and antiparticles.
 Thus, if the measurement of
 $\nu_{e}/\bar{\nu}_{e}$  ratio for high-energy  neutrinos originating
from AGNs were to become feasible, it may at least in principle
 constrain $\Delta f$
up to  (much) smaller values than which can currently be achieved
by
 neutrinos from other astrophysical sources \cite{Fogli}.
Let us note that the change in $\nu_{e}/\bar{\nu}_{e}$ ratio is not expected
from pure vacuum flavor oscillations. This can be a characteristic
 observational
consequence of incorporating the effect of
possible (uniform) rotation of magnetic strength lines along the high-energy
neutrino trajectories originating from AGNs.

    The expected event rates for different neutrino flavors in km$^{3}$
 volume high-energy neutrino telescopes using the rather optimistic diffuse
 upper flux limits, as an example,  given
in Ref. \cite{S} typically range, for $E\, \sim \, 10^{6}$ GeV, as follows:
 the downward going
 ($\nu_{e}+\bar{\nu}_{e}$) event rate is typically $\sim O(10^{1.5})$,
the downward going ($\nu_{\mu}+\bar{\nu}_{\mu}$) event rate is typically
$\sim O(10^{2})$, whereas the downward going ($\nu_{\tau}+\bar{\nu}_{\tau}$)
event rate is typically $\sim O(10^{1})$, all in units of per year per
steradian, the  downward going $\bar{\nu}_{e}$ event rate for
$E\sim 6.4 \cdot 10^{6}$ GeV
 being approximately half an order of magnitude higher than the
 ($\nu_{e}+\bar{\nu}_{e}$) event rate in the high-energy neutrino
 telescopes \cite{reno}.
 The three flavors are expected to have different event
topologies \cite{new}, thus providing some prospects to search for the
observational consequences pointed out in this section.

Summarizing,
a possible observational consequence of neutrino spin-flip in the high-energy
 neutrino telescopes include  a {\em change} in the
expected $\nu_{e}/\bar{\nu}_{e}$ ratio correlated to the
direction of source with an energy resolution
 $\Delta E/E\, {\buildrel < \over {_{\sim }}}\, 1/20$ near the Glashow
 resonance energy as well as a possible
 energy dependence in the ratio of the three flavors. Some  of the
other situations in neutrino spin-flip discussed here tend to overlap with
 the pure vacuum flavor oscillations scenario.

\section{Results and Discussion}

    The intrinsic fluxes of the high-energy neutrinos
 ($E\geq 10^{6}$ GeV)
originating from AGN are estimated
to have the following ratios:
 $(\nu_{e}+\bar{\nu}_{e})/(\nu_{\mu}+\bar{\nu}_{\mu})\, \simeq 1/2$, $
 (\nu_{\tau}+\bar{\nu}_{\tau})/(\nu_{e,\mu}+\bar{\nu}_{e,\mu})\,
 {\buildrel < \over {_{\sim}}} 10^{-5} $.
 Thus,
if an enhanced energy dependent
$(\nu_{\tau}+\bar{\nu}_{\tau})/(\nu_{e}+\bar{\nu}_{e})$ ratio
(as compared to no precession/conversion situation) is
observed correlated to the direction
of source for high-energy
neutrinos, then it may be either an evidence for a  spin-flip through
spin (flavor)-precession
 alone or through
 a resonant conversion in the vicinity of AGN due to
an interplay of VEP and/or a conceivable magnetic field twist
 in $B_{AGN}$ depending on the finer details of the relevant high-energy
 AGN  neutrino spectra.
    The spin (flavor)-precession and/or conversion effects
discussed in this paper  may be
distinguished from the pure vacuum flavor oscillations by observing
the {\em energy dependence} of the high-energy neutrino flux profiles.
 A mutual comparison of the relevant [that is, for instance,
 $(\nu_{e}+\bar{\nu}_{e})$ and
 $(\nu_{\tau}+\bar{\nu}_{\tau})$]  high-energy
 neutrino spectra may in principle isolate the mechanism of neutrino
conversions in the vicinity of AGN.

    The incorporation of a possible magnetic field twist
induces a level crossing
in the vicinity of AGN due to VEP.
 This conversion can be made adiabatically resonant for a naturally
scaled magnetic field twist  with
$\Delta f\, {\buildrel < \over {_{\sim}}} \, 10^{-34}$. A resonant
character in the
oscillations of high-energy neutrinos originating  from AGN
 for vanishing gravity and vacuum mixings may not be induced
otherwise. Thus, a breakdown in the universality of
gravitational coupling
of neutrinos  at the level of $10^{-34}$ or less depending on relevant
 $\delta m^{2}$ may provide a possible
cause for observing energy dependence and change in the
 three neutrino flavors w.r.t. pure vacuum flavor oscillations,
assuming that there is no appreciable reverse neutrino spin-flip
 between AGN and the earth.

    For small
$\delta m^{2}$ ($\delta m^{2}\, <\, 5\cdot 10^{-6}$ eV$^{2}$)
a  spin
(flavor)-precession may result in an energy
independent permutation of the relevant neutrino spectra with the
corresponding spin (flavor)-precession
probability greater than 1/2. This spin (flavor)-precession may
 occur for
small
$\Delta f$ ($\Delta f\, {\buildrel < \over {_{\sim}}} \, 10^{-34}$).
 The spin-flip may occur through resonant conversions induced by the
VEP and/or field twist in $B_{AGN}$ as
well. Assuming that the information on $\Delta f$ may be obtained from
various terrestrial/extraterrestrial experiments, a mutual comparison
 between
the survived and transformed high-energy AGN neutrinos may enable one
 to distinguish the mechanism of
conversion. If for small
$\delta m^{2}$ ($\delta m^{2}\, <\,
5\cdot 10^{-6}$ eV$^{2}$), an energy dependent permutation  are
obtained empirically with corresponding $P\, >\, 1/2$ then this
 situation may
 be an evidence for a conversion effect
due to an interplay of VEP and twist in $B_{AGN}$.

    For large
$\delta m^{2}$ ($\delta m^{2}\, >\, 5\cdot 10^{-6}$ eV$^{2}$),
if  energy dependent
distortions  and for instance a change in
 $(\nu_{\tau}+\bar{\nu}_{\tau})/(\nu_{e}+\bar{\nu}_{e}$) is observed
 with
the corresponding conversion
probability greater than 1/2 then the cause may be a relatively large
$\Delta f$ ($\Delta f\, >\, 10^{-34}$) and/or a naturally scaled magnetic
field twist. The level crossing induced by  VEP and/or field twist
 has a
{\em different} $E$
dependence thus in principle
with the improved information on either $\Delta f$ or the scale of
 magnetic
field twist, the cause of the
conversion effect may be isolated.
    Further, as the energy span in the relevant high-energy
 AGN neutrino
spectra is expected to be several  orders of
magnitude,
therefore, energy dependent spin (flavor)-precession/conversion
 probabilities
may result in distortions in some part(s) of the spectra for
relevant neutrino species and may thus be identifiable in future
high-energy neutrino telescopes.

    A possible observational consequence of neutrino spin-flip in AGN
in electron neutrino channel only can be
an observed {\em change} in $\nu_{e}/\bar{\nu}_{e}$ ratio
 (as compared to no spin-flip situation) near the Glashow resonance energy
 which
 may be  a result of an interplay of VEP and the magnetic field twist.
 This feature is absent in pure vacuum flavor oscillations.

    An additional feature of the present study is that it may
 provide
useful information on the strength/profile of $B_{AGN}$ if the cause of
$\nu_{e}\leftrightarrow \bar{\nu}_{\mu,\tau}$
conversion/precession can be  established due to
VEP and/or magnetic field twist for high-energy AGN neutrinos. \\

\paragraph*{Acknowledgments.}
The authors thank World Laboratory (WL) for financial support within project
E-13. HA also thanks Physics Division of National Center for Theoretical
Sciences for financial support.

\pagebreak

\begin{figure}[t]
\begin{center}
\leavevmode \epsfxsize=3.5in \epsfysize=3.5in
\epsfbox{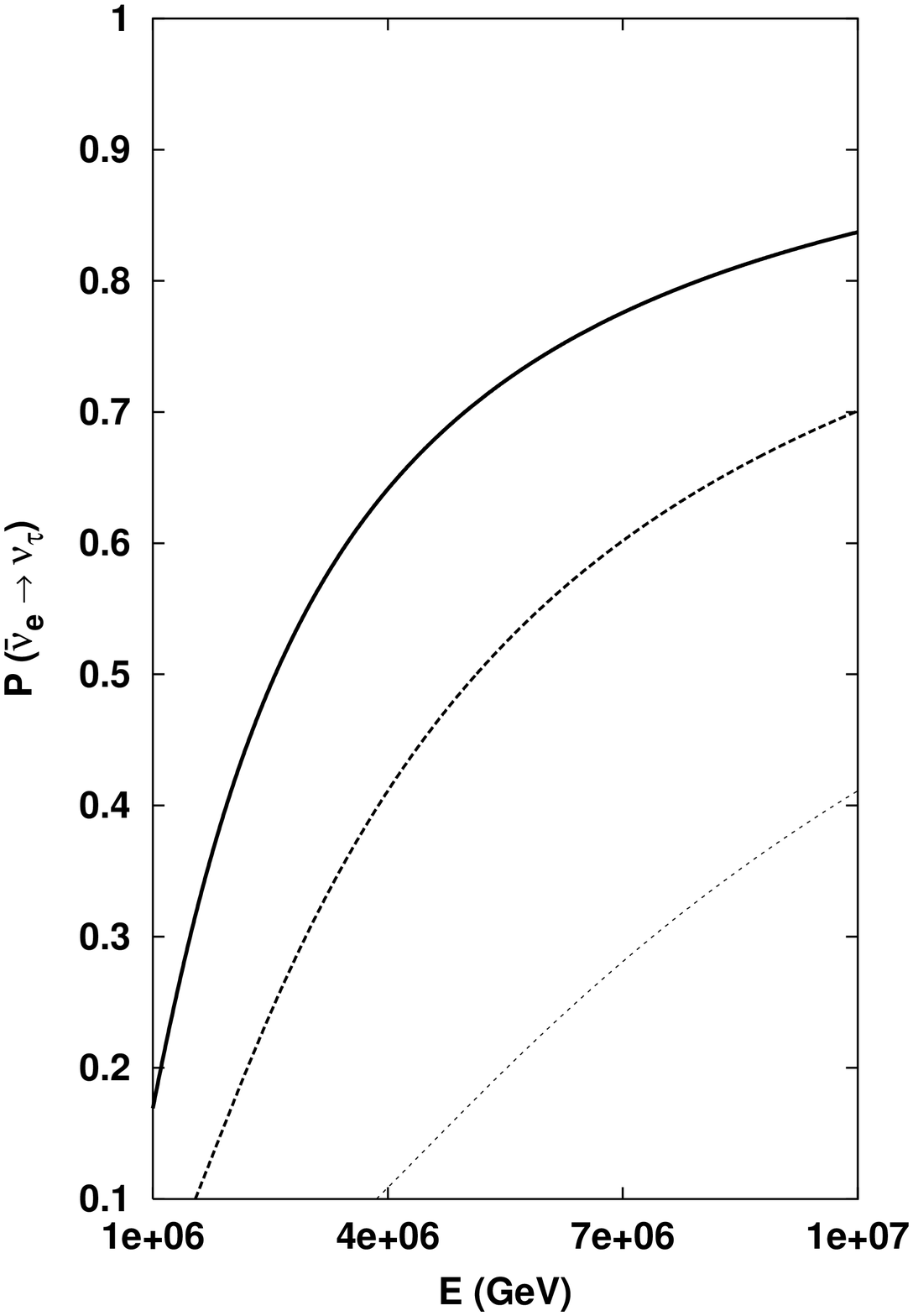} \vspace{1cm}
 \tightenlines
\caption{$P(\bar{\nu}_{e}\to \nu_{\tau})$ as a function of $E$
(GeV) for some
 representative values of $\Delta f $ with
 $\delta m^{2}\, \sim 10^{-10}$ eV$^{2}$ and $\mu \, \sim 10^{-12}\mu_{B}$
 using Eq. (9) for illustrative purpose. Upper curve, $\Delta f \,
 \sim 10^{-29}$,  middle curve, $\Delta f \,
 \sim 5 \cdot 10^{-30}$, lower curve, $\Delta f \,
 \sim 2 \cdot 10^{-30}$.}
\end{center}
\end{figure}


\begin{references}
\tightenlines
\bibitem{ss} See, for instance, F. W. Stecker, C. Done, M. H. Salamon, and
             P. Sommers, Phys. Rev. Lett. {\bf 66}, 2697 (1991);
             {\bf 69}, 2738 (E) 1992.
             For a recent update, see, R. Gandhi,
             Nucl. Phys. (Proc. Suppl.) {\bf B91},
             453 (2001).
\bibitem{G} M. Gasperini, Phys. Rev. {\bf  D38}, 2635 (1988); {\bf D39},
            3606 (1989).
\bibitem{L} For an independent similar possibility of testing the VEP by
            neutrinos, see, A. Halprin and C. N. Leung, Phys. Rev. Lett.
            {\bf  67}, 1833 (1991).
\bibitem{Fogli} For a recent discussion in the context of atmospheric muon
                neutrino deficit, see, for instance, G. L. Fogli
                {\em et al.}, Phys. Rev. {\bf D60}, 053006 (1999), and
                references cited therein, whereas for a recent discussion in
                the context of solar electron neutrino deficit, see, for
                instance,
                D. Majumdar, A. Raychaudhuri, and A. Sil,
                Phys. Rev. {\bf  D63}, 073014 (2001). For
                supernova neutrinos, see, M. M. Guzzo, H. Nunokawa, and
                R. Tom\`{a}s, hep-ph/0104054.
\bibitem{general} C. Aneziris and J. Schechter, Int. J. Mod. Phys. {\bf
                  A6}, 2375 (1991); S. Toshev, Phys. Lett. {\bf  B271},
                  179 (1991); C. Aneziris and J.  Schechter, Phys. Rev.
                  {\bf D45}, 1053 (1992); E. Kh. Akhmedov, S. T. Petcov and A.
                  Yu. Smirnov, Phys. Rev.  {\bf D48}, 2167 (1993);
                  {\em ibid}, Phys. Lett. {\bf  B309}, 95 (1993).
\bibitem{sun} J. Vidal and J. Wudka, Phys. Lett. {\bf B249}, 473 (1990);
              A. Yu. Smirnov, Phys. Lett. {\bf B260}, 161 (1991);
              E. Kh. Akhmedov, P. I. Krastev and A. Yu. Smirnov,
              Z. Phys.  {\bf C52}, 701 (1991);
              T. Kubota {\em et al}., Phys. Lett. {\bf  B292},
              195 (1992);
              M. Moretti,  Phys. Lett.  {\bf B293}, 378 (1992);
              M. M. Guzzo and J. Bellandi, Phys. Lett.  {\bf B294},
              243 (1992);
              P. I. Krastev, Phys. Lett.  {\bf B303}, 75 (1993);
              A. B. Balantekin and F. Loreti,  Phys. Rev.  {\bf D48},
              5496 (1993);
              M. M. Guzzo and J. Pulido, Phys. Lett. {\bf B317}, 125
              (1993);
              J. Bellandi and M. M. Guzzo,  Phys. Lett.  {\bf B317},
              130 (1993);
              T. Kubota {\em et al}.,  Phys. Rev.  {\bf D49}, 2462
              (1994);
              K. Ohta and E. Takasugi,  Prog. Theor. Phys. {\bf 92},
              733 (1994);
              V. M. Aquino, J. Bellandi and M. M. Guzzo,
              Physica Scripta {\bf  54}, 328 (1996);
              E. Kh. Akhmedov, hep-ph/9705451;
              J. Derkaoui and Y. Tayalati, Astropart. Phys. {\bf
              14}, 351 (2001).
\bibitem{supernova} H. Athar and J. T. Peltoniemi, Phys. Rev.
                    {\bf  D51}, 5785 (1995); H. Athar, J. T. Peltoniemi and
                    A. Yu. Smirnov, Phys. Rev. {\bf D51}, 6647 (1995).
\bibitem{universe} K. Enqvist and V. Semikoz, Phys. Lett.  {\bf B312},
                   310 (1993); V. Semikoz, Phys. Rev.  {\bf D48}, 5264
                   (1993); Erratum, {\em ibid}, {\bf D49}, 6246 (1994);
                   H. Athar, Phys. Lett. {\bf  B366}, 229 (1996).
\bibitem{shelly} Sheldon L. Glashow, Phys. Rev. {\bf 118}, 316 (1960).
\bibitem{john} J. G. Learned and S. Pakvasa, Astropart. Phys. {\bf 3},
               267 (1995).
\bibitem{lin} H. Athar and Guey-Lin Lin, hep-ph/0108204; {\em
              ibid}, hep-ph/0201026.
\bibitem{C} See, for a recent discussion on various high-energy neutrino
            telescopes, L. Moscoso,  Nucl. Phys.(Proc. Suppl.) {\bf B87},
            377 (2000). See also, C. Spiering,  Nucl. Phys.(Proc. Suppl.)
            {\bf B91}, 445 (2001); F. Halzen, astro-ph/0111059.
\bibitem{DKD} G. Domokos, S. Kovesi-Domokos, Phys. Lett. {\bf  B410},
              57 (1997); K. Enqvist, P. Ker\"{a}nen and J. Maalampi,
               Phys. Lett.  {\bf B438}, 295 (1998).
\bibitem{R} M. Roy, J. Phys.  {\bf G22}, L113 (1996).
\bibitem{sahu} Sarira Sahu and Vishnu M. Bannur, Mod. Phys. Lett.
               {\bf A15}, 775 (2000).
\bibitem{W} H. Athar and Jos\'{e} F. Nieves,  Phys. Rev. {\bf  D61},
            103001 (2000), and references cited therein.
\bibitem{V} M. B. Voloshin, M. I. Vysotoskii and L. B. Okun,  Sov. Phys.
            JETP {\bf  64}, 446 (1986).
\bibitem{luna} C. Lunardini and A. Yu. Smirnov,  Nucl. Phys. {\bf B583},
               260 (2000).
\bibitem{book} See, for instance, Edward W. Kolb and Michael S. Turner,
               {\em The Early Universe}
               (Addison-Wesely Publishing Company, Reading, Massachusetts,
               1990), pp. 244.
\bibitem{S} A. P. Szabo and R. J. Protheroe, Astropart. Phys. {\bf  2},
            375 (1994).
\bibitem{Ak} E. Kh. Akhmedov, S. T. Petcov and A. Yu. Smirnov, in Ref.
             \cite{general}.
\bibitem{japan} H. Minakata and A. Yu. Smirnov, Phys. Rev. {\bf D54},
                3698 (1996).
\bibitem{MPR} See, for instance, Ref. \cite{W} for a brief description of
              high-energy tau neutrino production, whereas for a recent
              discussion on high-energy non tau neutrino production,
              see, Karl Mannheim, R. J. Protheroe and J\"{o}rg  P. Rachen,
              Phys. Rev. {\bf  D63}, 023003 (2001), and
              references cited therein.
\bibitem{nu}  For a recent update on AMANDA, see, E. Andres {\em et al}.,
               Nucl. Phys.(Proc. Suppl.) {\bf B91},
              423 (2001).
\bibitem{wb} E. Waxman and J. N. Bahcall, Phys. Rev. {\bf D59},
                023002 (1999); Phys. Rev. {\bf D64}, 023002 (2001).
\bibitem{F} M. Fukugita and S. Yazaki, Phys. Rev. {\bf  D36}, 3817
            (1987); G. G. Raffelt, Astrophys. J.  {\bf 365}, 559 (1990);
            V. Castellani and S.  Degl'Innocenti, {\em ibid}, {\bf 402}, 574
            (1993) [the present upper bound on neutrino magnetic moment from
            laboratory experiments is, $\mu\, <\, 1.5 \cdot 10^{-10}\, \mu_{B}
            $, see, D. E. Groom {\em et al.}, Euro. Phys. J. {\bf C15},
            25 (2000)].
\bibitem{B} M. C. Begelman, R. G. Blandford and M. Rees,  Rev. Mod. Phys.
            {\bf  56}, 255 (1984).
\bibitem{YU} See, for instance, H. Sobel,  Nucl. Phys. (Proc. Suppl.)
             {\bf B91}, 127 (2001).
\bibitem{AJY} For a recent discussion, see, H. Athar, M. Je\.{z}abek and
              O. Yasuda, Phys. Rev.
              {\bf D62}, 103007 (2000), and references cited therein; H. Athar,
              hep-ph/0008121.
\bibitem{Sm} A. Yu. Smirnov, in Ref. \cite{sun}.
\bibitem{lim} C.-S. Lim and W. J. Marciano,  Phys. Rev. {\bf D37}, 1368
              (1988); E. Kh. Akhmedov, Sov. J. Nucl. Phys. {\bf 48},
              382 (1988) [ Yad. Fiz. {\bf 48}, 599 (1988)];
               Phys. Lett. {\bf B213}, 64 (1988).
\bibitem{kim} See, for instance, C. W. Kim and A. Pevsner,
              {\em Neutrinos in Physics and
              Astrophysics} (Harwood academic publishers, 1993), p. 264.
\bibitem{GH} Thomas K. Gaisser, Francis Halzen and Todor Stanev, Phys.
             Rep. {\bf 258}, 173 (1995) and erratum {\bf 271}, 355 (1996).
\bibitem{debate} Karl Mannheim,
                  J. Phys.  {\bf G27}, 1691 (2001).
\bibitem{global} See, for instance, M.~C.~Gonzalez-Garcia, M.~Maltoni,
                 C.~Pena-Garay and J.~W.~Valle,
                 Phys.\ Rev.\ {\bf D63}, 033005 (2001).
\bibitem{bona} H.~Athar, Astropart.\ Phys.\  {\bf 14}, 217 (2000).
\bibitem{aps} E. Kh. Akhmedov, S. T. Petcov and A. Yu. Smirnov, in
             \cite{general}.
\bibitem{reno} R. Gandhi, C. Quigg, M. H. Reno, and I. Sarcevic,
               Astropart. Phys. {\bf 5}, 81 (1996); Phys. Rev. {\bf
               D58}, 093009 (1998).
\bibitem{new} H. Athar, G. Parente,
               and E. Zas,  Phys. Rev. {\bf D62}, 093010 (2000), and
               references cited therein.

\end{references}
\end{document}